\documentclass[12pt,preprint]{aastex}
\usepackage{epstopdf}
\usepackage{natbib}

\newcommand{\peryear}{yr$^{-1}$\,}
\newcommand{\mjup}{~$M_{Jup}$\,}
\newcommand{\msun}{~$M_\odot$\,}

\begin{document}

\title{The Ultra Cool Brown Dwarf Companion of WD~0806--661: Age, Mass, and Formation Mechanism} 

\author{David R. Rodriguez\altaffilmark{1,2}, B.\ Zuckerman\altaffilmark{1}, Carl Melis\altaffilmark{2,3} \& Inseok Song\altaffilmark{4}}

\altaffiltext{1}{Dept.\ of Physics \& Astronomy, University of California, Los Angeles 90095, USA \\
(drodrigu@astro.ucla.edu)}
\altaffiltext{2}{Visiting astronomer, Cerro Tololo Inter-American Observatory, National Optical Astronomy Observatory, which are operated by the Association of Universities for Research in Astronomy, under contract with the National Science Foundation.}
\altaffiltext{3}{Center for Astrophysics and Space Sciences, University of California, San Diego, CA 92093, USA}
\altaffiltext{4}{Department of Physics and Astronomy, University of Georgia, Athens, GA 30602, USA}

\begin{abstract}

We have combined multi-epoch images from the Infrared Side Port Imager on the CTIO 4-meter telescope to derive a 3-$\sigma$ limit of $J=21.7$ for the ultra cool brown dwarf companion to WD~0806--661 (GJ~3483).
We find that $J-[4.5]>4.95$, redder than any other brown dwarf known to date.
With theoretical evolutionary models and ages $1.5-2.7$~Gyr, we estimate the brown dwarf companion to have mass $<10-13$\mjup and temperature $\lesssim400$~K, providing evidence that this is among the coolest brown dwarfs currently known.
The range of masses for this object is consistent with that anticipated from Jeans-mass fragmentation and we present this as the likely formation mechanism.
However, we find that substellar companions of similar mass ($\sim7-17$\mjup) are distributed over a wide range of semi-major axes, which suggests that giant planet and low-mass brown dwarf formation overlap in this mass range.

\end{abstract} 


\section{Introduction}

The first examples of the low-temperature L- and T-spectral classes were
discovered as companions to stars \citep{Becklin:1988,Nakajima:1995}.
The upper temperature limit for a Y-type object is not known; though it is expected to be $<500$~K, when water vapor is predicted to condense 
(see Figure~2 in \citealt{Burrows:2003}).
Wide-field searches for low
temperature T (and Y) dwarfs are currently in progress and several very cool objects have been found at temperatures of: 
$\sim$570~K \citep{Burningham:2009}, $550-500$~K \citep{Leggett:2009}, $500-400$~K \citep{Lucas:2010}, and $<$500~K \citep{Eisenhardt:2010}.
However, the spectra of those objects (when available) do not 
differ substantially from late T-dwarfs.
An object comoving with WD~0806--661 could represent the coolest substellar companion observed to date \citep{Luhman:2011}.
While no spectrum has yet been obtained, this could very well be the first example of the Y spectral class.

This Letter presents an upper limit at J-band to the substellar companion to WD~0806--661 (GJ~3483).
Because the `WD' designation is used exclusively for white dwarfs (E. Sion 2011, private comm.), we suggest the companion be referred to as GJ~3483B rather than WD~0806--661B, as adopted in the discovery paper \citep{Luhman:2011}.
We will use GJ~3483 and GJ~3483B throughout this Letter.
We discuss what constraints our upper limit places on the object's mass and temperature as well as how it ties in to other imaged low-mass companions.
This object, and similar low-mass companions, probably formed at large separations from their primary star via Jeans-mass fragmentation \citep{Low:1976} rather than been scattered out via planet-planet interactions.

\section{Observations}\label{observations}

As part of an ongoing imaging campaign to search for wide separation low-mass companions to nearby, old white dwarfs, we observed GJ~3483 using the Infrared Side Port Imager (ISPI, \citealt{van-der-Bliek:2004}) on the CTIO 4-m Blanco telescope.
Observations were carried out on UT 2009-03-16, 2009-03-17, 2010-04-03, and 2010-04-05, each lasting 3240 seconds (3 sets of $3\times3$ dithers with individual exposure times of 120 seconds) and covering a field of view of approximately $10\arcmin \times 10\arcmin$.
Conditions were generally good with seeing $\sim$1\arcsec~at J-band ($\sim$1.3\arcsec~on 2009-03-17) and clear (except for thin clouds on 2009-03-16).
Data reduction was carried out in the usual manner using standard IRAF routines\footnote{IRAF is distributed by the National Optical Astronomy Observatory, which is operated by the Association of Universities for Research in Astronomy (AURA) under cooperative agreement with the National Science Foundation.}.
The brown dwarf companion observed by Spitzer \citep{Luhman:2011} was not detected in any of the individual nights.

\begin{figure}[htb]
\begin{center}
\includegraphics[width=14cm,angle=0]{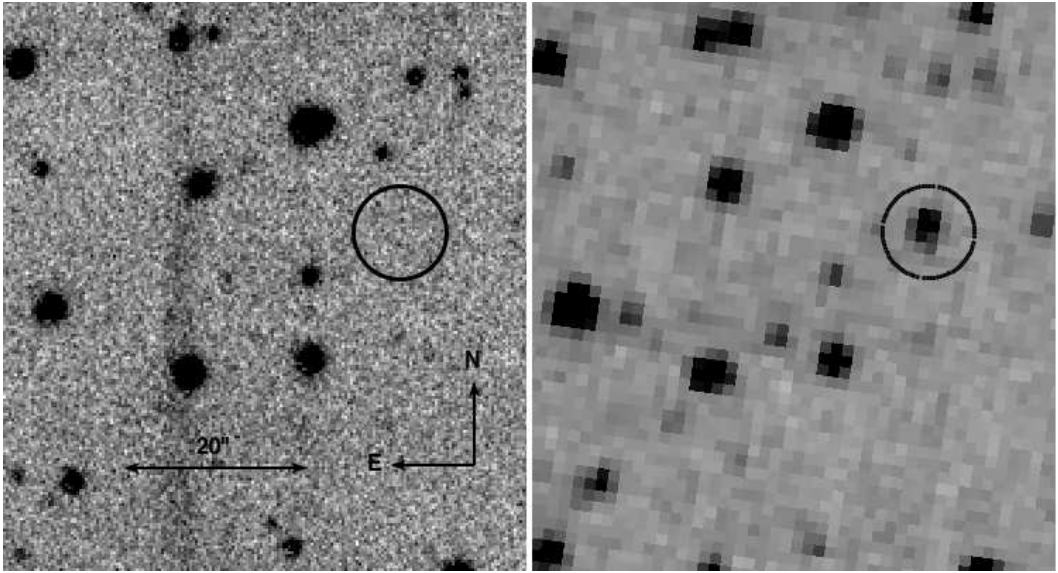}
\end{center}
\caption{
Co-added (3.5 hours) CTIO ISPI J-band field (left) and Spitzer IRAC 4.5$\mu$m field (right); the circle denotes the location of GJ~3483B.
}
\label{fig:image}
\end{figure}

To obtain a deeper J-band limit, we coadded data from the four nights together.
Because the white dwarf has proper motion of ($+340.3$, $-289.6$) mas~\peryear \citep{Subasavage:2009}, we shifted the 2009 data by 1~pixel in both R.A.\ and Decl.\ to match the white dwarf's motion (the ISPI plate scale is 0.3\arcsec~pix$^{-1}$). 
Individual nights were weighted by the square of the signal-to-noise of well-detected 2MASS stars.
Our coadded, 3.5-hour image is displayed in Figure~\ref{fig:image} alongside the 2009 $4.5\mu$m Spitzer field from program 60160 (M.\ Burleigh) and it is clear we do not detect the companion brown dwarf.
To estimate a limiting magnitude, we generated synthetic stars in 0.1 magnitude bins ranging from J=20.6 to 22.6 using IRAF/DAOPHOT's {\sf addstar} routine with a PSF derived from stars in our field.
PSF fitting was performed on these synthetic stars, which allows us to estimate a 3-$\sigma$ limit of $J=21.7$ for this coadded exposure, 1.7~magnitudes fainter than the limit provided by \citet{Luhman:2011}.

\section{Results}\label{resultsWD}

\begin{figure}[htb]
\begin{center}
\includegraphics[width=14cm,angle=0]{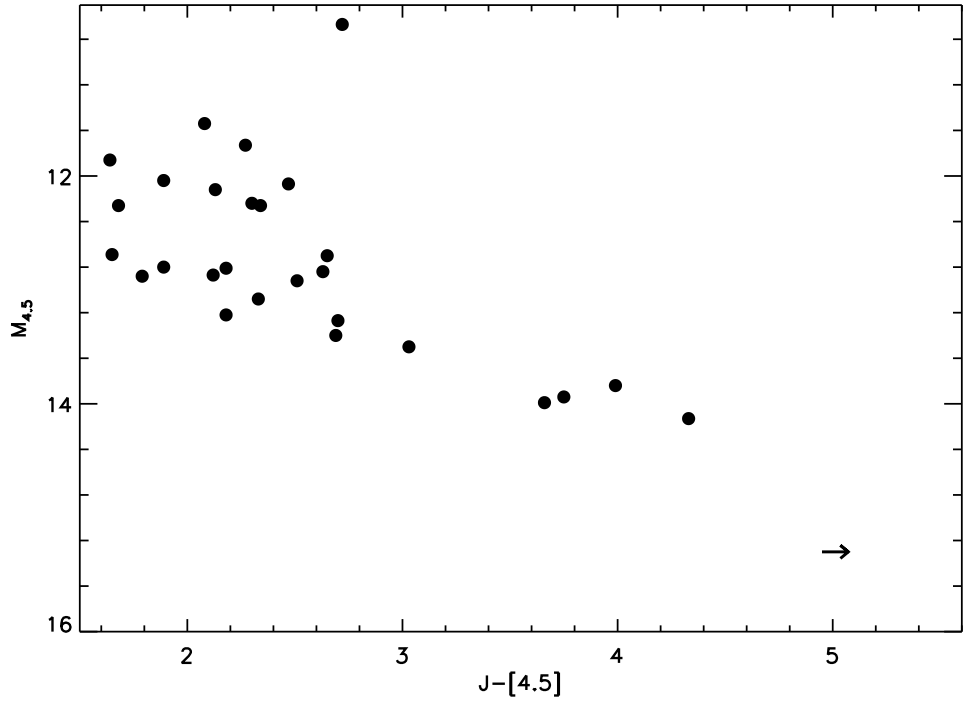}
\end{center}
\caption{
$M_{4.5}$ versus $J-[4.5]$ for T dwarfs with measured distances and IRAC photometry from \citet{Lucas:2010} and \citet{Leggett:2010}. GJ~3483B is indicated with a right-pointing arrow whose base corresponds to our upper limit of $J-[4.5]=4.95$. It is the reddest brown dwarf currently known.
}
\label{fig:m45}
\end{figure}

With our upper limit of $J=21.7$, we estimate $J-[4.5]>4.95$, nearly two magnitudes greater than the value quoted in \citet{Luhman:2011} and redder than any other T-dwarf currently known.
Figure~\ref{fig:m45} illustrates the trend of decreasing absolute $4.5\mu$m magnitude for known T-dwarfs as a function of $J-[4.5]$ color. Our upper limit for GJ~3483B is consistent with the T-dwarf trend, though it could be redder (and, thus, as faint as $J\sim23-24$).
The initial-final mass relationships in \citet[and references therein]{Catalan:2008} along with the estimated white dwarf mass of $0.62\pm0.03$\msun \citep{Subasavage:2009} suggest the progenitor mass was $2.0\pm0.3$\msun with a main sequence lifetime of $1.3_{-0.5}^{+0.7}$~Gyr (\citealt{Catalan:2008}, and references therein; S.~Leggett 2011, private comm.). 
With the cooling age of 670~Myr derived in \citet{Subasavage:2009}, this implies a total age of 1.5 to 2.7~Gyr.
Based on our 3-$\sigma$ limit, an age of 1.5~Gyr, and theoretical evolutionary models from \citet{Burrows:2003} and \citet{Baraffe:2003}, we estimate a mass of $<10$\mjup and temperature $<410$~K.
However, with an age of 2.7~Gyr, we find the companion has a mass of $<13$\mjup and temperature $<400$~K.

The IRAC 4.5$\mu$m photometric magnitude of 16.72 \citep{Luhman:2011} can also be used to derive the mass and temperature of GJ~3483B.
With an age of 1.5~Gyr, we find the \citet{Burrows:2003} models predict a mass of $\sim6$\mjup and temperature $\sim330$~K.
With the upper age limit of 2.7~Gyr, we find mass and temperature of $\sim10$\mjup and $\sim350$~K.
These model temperatures are somewhat warmer than the value \citet{Luhman:2011} present ($\sim300$~K).
While the uncertainty in age and thus the mass of GJ~3483B are substantial, the range of implied masses is consistent with those anticipated from Jeans-mass fragmentation (see Section~\ref{discussionWD}). 
That is, the mass of GJ~3483B is unlikely to be much smaller than $\sim7$\mjup.
The range in mass and temperature ($6-10$\mjup, $330-350$~K) for GJ~3483B are comparable to those of CFBDSIR~J1458+1013B ($6-15$\mjup, $370\pm40$~K), another recently discovered ultra cool companion \citep{Liu:2011}.

\section{Discussion}\label{discussionWD}

Companions with masses less than $\sim$20\mjup~have been imaged in only a handful of systems (listed in Table~\ref{tab:lowmass}) and most have masses between 7 and 15\mjup.
As described in \citet{Zuckerman:2009}, objects of mass $\sim$15~M$_{Jup}$
would require at least 2~Gyr to cool down to $\sim$500~K, which is usually taken as the temperature in which the Y-dwarf spectral type will appear.
Binaries with brown dwarf secondaries having masses larger than 15 M$_{Jup}$
are somewhat more common than the few listed in Table~\ref{tab:lowmass} (see
Table 2 in \citealt{Zuckerman:2009}), 
but for a $\sim$25~M$_{Jup}$ object to cool to 500~K it must be at least
7~Gyr old, according to the \citet{Burrows:2003} models.
Hence the coolest companions are more likely to be found in relatively old systems.

A handful of white dwarfs are known to host widely separated brown dwarf companions (see Table~\ref{tab:wdbd}). 
While a few other WD+BD pairs are known, these are unresolved systems whose separations are a few AU or less (GD 1400, \citealt{Farihi:2004}; WD~0137-349, \citealt{Maxted:2006}).
The separations among the resolved WD+BD pairs is generally on the order of $\sim$1000's of AU, albeit with low number statistics.
Companions around white dwarfs are expected (and observed, see \citealt{Farihi:2010}) to have a bimodal distribution in their orbital semi-major axes \citep{Nordhaus:2010}.
Those orbiting close to the main sequence star will spiral in via tidal dissipation \citep{Hansen:2010}.
More widely separated companions (such as most of those in Table~\ref{tab:lowmass}) will 
undergo orbital expansion following post main sequence
mass loss from the white dwarf progenitor. Assuming the mass
loss was adiabatic, which is almost certainly true (the orbital period is
short compared to the time frame of mass-loss), the ratio
of semimajor axes ($a$) between the white dwarf and main sequence
phases should be identical to the inverse ratio of white dwarf
and main sequence masses, or $a_{\rm wd}=a_{\rm ms}M_{\rm ms}
/M_{\rm wd}$ \citep{Jeans:1924}. 
As previously mentioned, GJ~3483 has a final mass of 0.62\msun~and initial mass of 2.0\msun.
If we assume the 2500~AU projected separation corresponds to a semi-major axis, then while on the main sequence this binary would have had $a\sim780$~AU, consistent with the values in Table~\ref{tab:lowmass}.

The commonly accepted mass limits for brown dwarfs are 13 and 75\mjup, where an object would fuse deuterium, but not hydrogen. In this case, planets (in addition to orbiting stars) would have masses below $\sim$13\mjup.
However, Jeans fragmentation can produce gravitationally collapsing clumps with masses as low as $\sim$7\mjup \citep{Low:1976}.
Typically, these fragments will have average separations of a few 100~AU.
As seen in Table~\ref{tab:lowmass}, there are several imaged systems with masses lower than $\sim$15\mjup~that orbit their host stars at these wide separations.
Giant planets, on the other hand, can form via core accretion in a disk fairly close to the star \citep{Pollack:1996,Kenyon:2009} or via gravitational instability at larger separations so long as the disk is sufficiently massive \citep{Boss:1997}.
Numerical simulations, however, have shown that planet formation via core accretion does not generally occur beyond 35~AU (\citealt{Dodson-Robinson:2009}, but see also \citealt{Currie:2011} for contrasting arguments).
Gravitational instability could create massive planets at larger semi-major axes, but only for the most massive disks would one expect planets to form at $\sim$100~AU (see \citealt{Rafikov:2005}, and references therein).

\begin{figure}[htb]
\begin{center}
\includegraphics[width=14cm,angle=0]{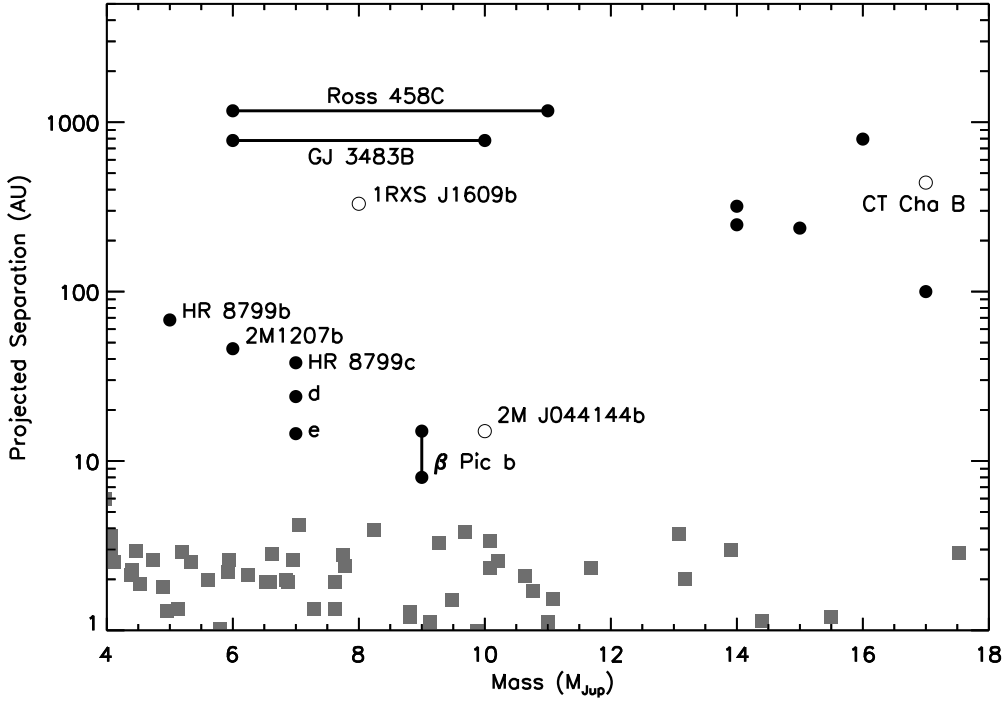}
\end{center}
\caption{
The lowest mass imaged companions to date (listed in Table~\ref{tab:lowmass}).
GJ~3483B is plotted at its expected location while the primary was on main sequence ($\sim780$~AU; see Section~\ref{discussionWD}); we also indicate the mass range ($6-10$\mjup). 
Gray squares denote extrasolar planet candidates from the Exoplanet Orbit Database \citep{Wright:2010}; plotted masses are M~sin$i$ and could be larger than shown.
CFBDSIR~J1458+1013B is not shown; its separation is 2.6~AU, but has a broad range for the mass ($\sim6-15$\mjup; \citealt{Liu:2011}).
CT~Cha~B, 1RXS~J1609b, and 2M~J044144b (whose mass may be 5--10\mjup, though for clarity we only display the higher value from \citealt{Todorov:2010}) are displayed as open circles as it's not yet known if these are bound (see footnote to Table~\ref{tab:lowmass}).
}
\label{fig:lowmass}
\end{figure}

Figure~\ref{fig:lowmass} shows the separations and masses of low mass companions from Table~\ref{tab:lowmass} and extrasolar planets from the Exoplanet Orbit Database \citep{Wright:2010}.
The population of exoplanets extends to lower masses and shorter separations (semi-major axes), but is not displayed here.
As mentioned in the preceding paragraphs, one would not expect giant planets to form at separations larger than $\sim100$~AU.
However, it appears that for objects $\sim7-17$\mjup, companions are distributed in a wide range of separations, reaching as far as $\sim1000$~AU. 
Planet-planet scattering could conceivably eject objects out to these large separations \citep[and references therein]{Veras:2009,Dodson-Robinson:2009}.
Since lower mass planets are more readily scattered out to larger orbits than more massive objects, one would expect to see a larger number of low mass planets at large separations.
However, numerous adaptive optics searches for companions have been performed, yet very few low mass objects have been found despite sensitivities to objects with masses $\gtrsim3$\mjup at separations $\gtrsim$40~AU (\citealt{Chauvin:2010,Nielsen:2010}; and references therein).
Additionally, while many giant planets have been detected with radial velocity searches, there is a lack of $>13$\mjup substellar companions within $\lesssim5$~AU, despite being sensitive to such objects (the well-known brown dwarf desert, see \citealt{Marcy:2000}).
The paucity of low mass companions at large separations contrasts with both theoretical expectations and detected giant planets at $a\lesssim5$~AU and suggests that planet-planet scattering is not a common mechanism to move large, $\sim7-17$\mjup mass companions, such as GJ~3483B and those listed in Table~\ref{tab:lowmass}, to large separations.

The low mass, directly imaged companions with separations $\lesssim100$~AU are worth mentioning, as they may be notable exceptions to Jeans fragmentation: HR~8799b,c,d,e; $\beta$~Pic~b; and 2M1207b.
In the case of HR~8799 and $\beta$~Pic, the planetary mass companions orbit relatively close to their A-type stars, which possess extended circumstellar disks (see \citealt{Rhee:2007}). 
The brown dwarf 2M1207 also possesses a disk close to the primary ($R\lesssim0.2$~AU; \citealt{Riaz:2006a}), but the low mass companion orbits beyond it at 46~AU.
While the disks around A-type stars can be massive enough to form giant planets at large separations, it is unlikely that a brown dwarf can possess a similarly massive disk.
This suggests that, while low mass brown dwarf formation and planet formation operate on different scales, formation mechanisms for companions with masses $\lesssim13$\mjup~begin to overlap 
(see discussions in \citealt{Currie:2011} and \citealt{Kratter:2010}).
The discovery of additional low mass, widely separated companions will play a key role in our understanding of giant planet and low mass brown dwarf formation.

\acknowledgements
{\it Acknowledgements.} 
We thank Adam Burgasser, Kevin Luhman, and Patrick Dufour for useful discussions.
We thank our referee, Sandy Leggett, for a prompt review and useful suggestions that strengthened this paper.
We would also like to thank Jay Farihi for bringing to our attention the incorrect use of WD designations for non-white dwarf companions to white dwarf stars.
This research has made use of the Exoplanet Orbit Database and the Exoplanet Data Explorer at exoplanets.org.
This research was supported by NASA Astrophysics Data Analysis Program grant NNX09AC96G to RIT and UCLA.

\clearpage

\begin{table}[htb!]
\caption{Lowest mass imaged companions}
\label{tab:lowmass}
\begin{center} {\scriptsize
\begin{tabular}{ccccccccl}
\hline
 Object & \multicolumn{2}{c}{Sp. Type} & Age & $M_{pri}$ & $M_{sec}$ & \multicolumn{2}{c}{Sep.} & Ref. \\
  & Primary & Sec. & (Myr) & (M$_{\odot}$) & ($M_{Jup}$) & (AU) & ($\arcsec$)   &  \\
\hline
\hline
2M1207--39  &   M8    &   L5    &$     8     $&  0.025 &  6  &  46 & 0.9 & 1   \\
AB Pic  &   K2V   &   L1    &$    30     $&  0.84  & 14  & 248 & 5.5 & 1 \\
Oph 11 &   M9    &   M9.5  &$     5     $&  0.0175& 15  & 237 & 1.6 & 1  \\
GQ Lup  &   K7V   &   L1.5  &$     3?    $&  0.7   & 17  & 100 & 0.7 &  1  \\
HN Peg &  G0V   &   T2.5  &$    200    $&  1.0   & 16  & 795 & 43.2 & 1  \\
HR 8799 	      &   A5V     &  late-L  &$ 30      $ & 1.5   & 5  &   68   & 1.7 & 2,3\\
         & 	      &	&	  &       & 7 &  38  & 1.0 & \\
         & 	      &	&	  &       & 7 &  24  & 0.6 & \\
	 &           &        &        &       & 7 &  14.5  & 0.4 & \\
$\beta$ Pic & A5V & late-L & 12 & 1.75 & $\sim$9 & 8--15 & 0.3 & 4\\	
Ross 458 & M0.5+M7 & T8 & $150-800$ & 0.6+0.08 & 6--11 & 1168 & 102 & 5,6\\             
GSC 06214--00210 & M1 & M8--L4 & 5 & 0.6 & 14 & 319 & 2.2 & 7\\
GJ~3483 & DQ & Y? & 1500--2700 & 0.62 & $\sim6-10$ & 2500$^{\bf a}$ & 130 & 8,9\\
\hline
\end{tabular} }
\end{center}
\footnotesize{$^{\bf a}$While on the main sequence, the orbital separation for GJ~3483B may have been $\sim$780~AU (see Section~\ref{discussionWD}).\\}
\footnotesize{{\bf References:} 1--~\citet{Zuckerman:2009}; and references therein, 2-- \citet{Marois:2008}, 3-- \citet{Marois:2010}, 4-- \citet{Lagrange:2010}, 5--~\citet{Goldman:2010}, 6--~\citet{Burgasser:2010}, 7-- \citet{Ireland:2011}, 8-- \citet{Luhman:2011}, 9--~This work. \\}
\footnotesize{ 
We do not list the recently announced 2.6~AU-separation CFBDSIR J1458+1013AB system as it lacks a firm age estimate and thus has widely varying masses for the primary and secondary, both of which are brown dwarfs with likely masses $<20$\mjup \citep{Liu:2011}.
A few other low-mass companions are known, but either with uncertain masses that could well be above $\sim$20\mjup (for example, GJ 758B at 10--40\mjup; \citealt{Thalmann:2009}) or the possibility that the two components could be independent members of the same co-moving region 
(such as 2M~J04414489+2301513, 1RXS~J160929.1--210524, and CT~Cha; \citealt{Todorov:2010,Lafreniere:2010,Schmidt:2008}).
The measured separations are lower limits to the semi-major axes which, on average, will be somewhat larger than the observed separations.
\\ }
\end{table}


\begin{table}[htb!]
\caption{Spatially resolved WD+BD pairs}
\label{tab:wdbd}
\begin{center} {\scriptsize
\begin{tabular}{cccccccccl}
\hline
 Object & \multicolumn{2}{c}{Sp. Type} & Age & $M_{pri}$ & $M_{sec}$ & $T_{eff}$ & \multicolumn{2}{c}{Sep.} & Ref. \\
  & Primary & Sec. & (Gyr) & (M$_{\odot}$) & ($M_{Jup}$) & (K) & (AU) & ($\arcsec$)   &  \\
\hline
\hline
GD~165 & DA & L3 & $1.2-5.5$ & $0.56-0.65$ & $\sim$75 & 1900 & 140 & 4 & 1,2\\
PHL~5038 & DA & L8 & $1.9-2.7$ & $0.57-0.87$ & 60 & 1450 & 55 & 0.94 & 3\\
LSPM~1459+0857AB & DA & T4.5 & $>4.8$ & 0.585 & $64-75$ & 1350 & $16500-26500$ & 365 & 4\\
GJ~3483 & DQ & Y? & $1.5-2.7$ & 0.62 & $\sim6-10$ & $340$ & 2500 & 130 & 5,6\\
 \hline
\end{tabular} }
\end{center}
\footnotesize{ {\bf References:} 1-- \citet{Zuckerman:1992}, 2-- \citet{Kirkpatrick:1999}, 3-- \citet{Steele:2009}, 4-- \citet{Day-Jones:2011}, 5-- \citet{Luhman:2011}, 6--~This work. \\}
\footnotesize{ Temperatures listed are averages of the range provided in the references (including GJ~3483B) for the secondary in the system. A range of separation is listed for LSPM~1459+0857AB as the distance is determined photometrically assuming the object is either single or an unresolved binary \citep{Day-Jones:2011}. Separations for these WD+BD pairs would have been smaller while on the main sequence, typically by a factor of 2 to 4 (see Section~\ref{discussionWD}).
}
\end{table}

\clearpage

\begin{thebibliography}{50}
\expandafter\ifx\csname natexlab\endcsname\relax\def\natexlab#1{#1}\fi

\bibitem[{Baraffe {et~al.}(2003)Baraffe, Chabrier, Barman, Allard, \&
  Hauschildt}]{Baraffe:2003}
Baraffe, I., Chabrier, G., Barman, T., Allard, F., \& Hauschildt, P.~H. 2003,
  \aap, 402, 701

\bibitem[{{Becklin} \& {Zuckerman}(1988)}]{Becklin:1988}
{Becklin}, E.~E., \& {Zuckerman}, B. 1988, \nat, 336, 656

\bibitem[{{Boss}(1997)}]{Boss:1997}
{Boss}, A.~P. 1997, Science, 276, 1836

\bibitem[{{Burgasser} {et~al.}(2010){Burgasser}, {Simcoe}, {Bochanski},
  {Saumon}, {Mamajek}, {Cushing}, {Marley}, {McMurtry}, {Pipher}, \&
  {Forrest}}]{Burgasser:2010}
{Burgasser}, A.~J., {et~al.} 2010, \apj, 725, 1405

\bibitem[{Burningham {et~al.}(2009)Burningham, Pinfield, Leggett, Tinney, Liu,
  Homeier, West, Day-Jones, Huelamo, Dupuy, Zhang, Murray, Lodieu,
  Navascu{\'e}s, Folkes, Galvez-Ortiz, Jones, Lucas, Calderon, \&
  Tamura}]{Burningham:2009}
Burningham, B., {et~al.} 2009, \mnras, 395, 1237

\bibitem[{Burrows {et~al.}(2003)Burrows, Sudarsky, \& Lunine}]{Burrows:2003}
Burrows, A., Sudarsky, D., \& Lunine, J.~I. 2003, \apj, 596, 587

\bibitem[{{Catal{\'a}n} {et~al.}(2008){Catal{\'a}n}, {Isern},
  {Garc{\'{\i}}a-Berro}, \& {Ribas}}]{Catalan:2008}
{Catal{\'a}n}, S., {Isern}, J., {Garc{\'{\i}}a-Berro}, E., \& {Ribas}, I. 2008,
  \mnras, 387, 1693

\bibitem[{{Chauvin} {et~al.}(2010){Chauvin}, {Lagrange}, {Bonavita},
  {Zuckerman}, {Dumas}, {Bessell}, {Beuzit}, {Bonnefoy}, {Desidera}, {Farihi},
  {Lowrance}, {Mouillet}, \& {Song}}]{Chauvin:2010}
{Chauvin}, G., {et~al.} 2010, \aap, 509, A52

\bibitem[{{Currie} {et~al.}(2011){Currie}, {Burrows}, {Itoh}, {Matsumura},
  {Fukagawa}, {Apai}, {Madhusudhan}, {Hinz}, {Rodigas}, {Kasper}, {Pyo}, \&
  {Ogino}}]{Currie:2011}
{Currie}, T., {et~al.} 2011, \apj, 729, 128

\bibitem[{Day-Jones {et~al.}(2011)Day-Jones, Pinfield, Ruiz, Beaumont,
  Burningham, Gallardo, Gianninas, Bergeron, Napiwotzki, Jenkins, Zhang,
  Murray, Catal{\'a}n, \& Gomes}]{Day-Jones:2011}
Day-Jones, A.~C., {et~al.} 2011, \mnras, 410, 705

\bibitem[{{Dodson-Robinson} {et~al.}(2009){Dodson-Robinson}, {Veras}, {Ford},
  \& {Beichman}}]{Dodson-Robinson:2009}
{Dodson-Robinson}, S.~E., {Veras}, D., {Ford}, E.~B., \& {Beichman}, C.~A.
  2009, \apj, 707, 79

\bibitem[{{Eisenhardt} {et~al.}(2010){Eisenhardt}, {Griffith}, {Stern},
  {Wright}, {Ashby}, {Brodwin}, {Brown}, {Bussmann}, {Dey}, {Ghez}, {Glikman},
  {Gonzalez}, {Kirkpatrick}, {Konopacky}, {Mainzer}, {Vollbach}, \&
  {Wright}}]{Eisenhardt:2010}
{Eisenhardt}, P.~R.~M., {et~al.} 2010, \aj, 139, 2455

\bibitem[{{Farihi} \& {Christopher}(2004)}]{Farihi:2004}
{Farihi}, J., \& {Christopher}, M. 2004, \aj, 128, 1868

\bibitem[{{Farihi} {et~al.}(2010){Farihi}, {Hoard}, \& {Wachter}}]{Farihi:2010}
{Farihi}, J., {Hoard}, D.~W., \& {Wachter}, S. 2010, \apjs, 190, 275

\bibitem[{{Goldman} {et~al.}(2010){Goldman}, {Marsat}, {Henning}, {Clemens}, \&
  {Greiner}}]{Goldman:2010}
{Goldman}, B., {Marsat}, S., {Henning}, T., {Clemens}, C., \& {Greiner}, J.
  2010, \mnras, 405, 1140

\bibitem[{{Hansen}(2010)}]{Hansen:2010}
{Hansen}, B.~M.~S. 2010, \apj, 723, 285

\bibitem[{{Ireland} {et~al.}(2011){Ireland}, {Kraus}, {Martinache}, {Law}, \&
  {Hillenbrand}}]{Ireland:2011}
{Ireland}, M.~J., {Kraus}, A., {Martinache}, F., {Law}, N., \& {Hillenbrand},
  L.~A. 2011, \apj, 726, 113

\bibitem[{Jeans(1924)}]{Jeans:1924}
Jeans, J.~H. 1924, \mnras, 85, 2

\bibitem[{{Kenyon} \& {Bromley}(2009)}]{Kenyon:2009}
{Kenyon}, S.~J., \& {Bromley}, B.~C. 2009, \apjl, 690, L140

\bibitem[{Kirkpatrick {et~al.}(1999)Kirkpatrick, Reid, Liebert, Cutri, Nelson,
  Beichman, Dahn, Monet, Gizis, \& Skrutskie}]{Kirkpatrick:1999}
Kirkpatrick, J.~D., {et~al.} 1999, \apj, 519, 802

\bibitem[{{Kratter} {et~al.}(2010){Kratter}, {Murray-Clay}, \&
  {Youdin}}]{Kratter:2010}
{Kratter}, K.~M., {Murray-Clay}, R.~A., \& {Youdin}, A.~N. 2010, \apj, 710,
  1375

\bibitem[{Lafreni{\`e}re {et~al.}(2010)Lafreni{\`e}re, Jayawardhana, \& van
  Kerkwijk}]{Lafreniere:2010}
Lafreni{\`e}re, D., Jayawardhana, R., \& van Kerkwijk, M.~H. 2010, \apj, 719,
  497

\bibitem[{Lagrange {et~al.}(2010)Lagrange, Bonnefoy, Chauvin, Apai, Ehrenreich,
  Boccaletti, Gratadour, Rouan, Mouillet, Lacour, \& Kasper}]{Lagrange:2010}
Lagrange, A.-M., {et~al.} 2010, Science, 329, 57

\bibitem[{Leggett {et~al.}(2009)Leggett, Cushing, Saumon, Marley, Roellig,
  Warren, Burningham, Jones, Kirkpatrick, Lodieu, Lucas, Mainzer, Mart{\'\i}n,
  McCaughrean, Pinfield, Sloan, Smart, Tamura, \& Cleve}]{Leggett:2009}
Leggett, S.~K., {et~al.} 2009, \apj, 695, 1517

\bibitem[{Leggett {et~al.}(2010)Leggett, Burningham, Saumon, Marley, Warren,
  Smart, Jones, Lucas, Pinfield, \& Tamura}]{Leggett:2010}
---. 2010, \apj, 710, 1627

\bibitem[{{Liu} {et~al.}(2011){Liu}, {Delorme}, {Dupuy}, {Bowler}, {Albert},
  {Artigau}, {Reyle}, {Forveille}, \& {Delfosse}}]{Liu:2011}
{Liu}, M.~C., {et~al.} 2011, arXiv:1103.0014

\bibitem[{Low \& Lynden-Bell(1976)}]{Low:1976}
Low, C., \& Lynden-Bell, D. 1976, \mnras, 176, 367

\bibitem[{{Lucas} {et~al.}(2010){Lucas}, {Tinney}, {Burningham}, {Leggett},
  {Pinfield}, {Smart}, {Jones}, {Marocco}, {Barber}, {Yurchenko}, {Tennyson},
  {Ishii}, {Tamura}, {Day-Jones}, {Adamson}, {Allard}, \&
  {Homeier}}]{Lucas:2010}
{Lucas}, P.~W., {et~al.} 2010, \mnras, 408, L56

\bibitem[{{Luhman} {et~al.}(2011){Luhman}, {Burgasser}, \&
  {Bochansky}}]{Luhman:2011}
{Luhman}, K.~L., {Burgasser}, A.~J., \& {Bochansky}, J.~J. 2011,
  arXiv:1102.5411

\bibitem[{Marcy \& Butler(2000)}]{Marcy:2000}
Marcy, G.~W., \& Butler, R.~P. 2000, Publications of the Astronomical Society
  of the Pacific, 112, 137

\bibitem[{Marois {et~al.}(2008)Marois, Macintosh, Barman, Zuckerman, Song,
  Patience, Lafreni{\`e}re, \& Doyon}]{Marois:2008}
Marois, C., Macintosh, B., Barman, T., Zuckerman, B., Song, I., Patience, J.,
  Lafreni{\`e}re, D., \& Doyon, R. 2008, Science, 322, 1348

\bibitem[{Marois {et~al.}(2010)Marois, Zuckerman, Konopacky, Macintosh, \&
  Barman}]{Marois:2010}
Marois, C., Zuckerman, B., Konopacky, Q.~M., Macintosh, B., \& Barman, T. 2010,
  Nature, 468, 1080

\bibitem[{{Maxted} {et~al.}(2006){Maxted}, {Napiwotzki}, {Dobbie}, \&
  {Burleigh}}]{Maxted:2006}
{Maxted}, P.~F.~L., {Napiwotzki}, R., {Dobbie}, P.~D., \& {Burleigh}, M.~R.
  2006, \nat, 442, 543

\bibitem[{{Nakajima} {et~al.}(1995){Nakajima}, {Oppenheimer}, {Kulkarni},
  {Golimowski}, {Matthews}, \& {Durrance}}]{Nakajima:1995}
{Nakajima}, T., {Oppenheimer}, B.~R., {Kulkarni}, S.~R., {Golimowski}, D.~A.,
  {Matthews}, K., \& {Durrance}, S.~T. 1995, \nat, 378, 463

\bibitem[{{Nielsen} \& {Close}(2010)}]{Nielsen:2010}
{Nielsen}, E.~L., \& {Close}, L.~M. 2010, \apj, 717, 878

\bibitem[{{Nordhaus} {et~al.}(2010){Nordhaus}, {Spiegel}, {Ibgui}, {Goodman},
  \& {Burrows}}]{Nordhaus:2010}
{Nordhaus}, J., {Spiegel}, D.~S., {Ibgui}, L., {Goodman}, J., \& {Burrows}, A.
  2010, \mnras, 408, 631

\bibitem[{{Pollack} {et~al.}(1996){Pollack}, {Hubickyj}, {Bodenheimer},
  {Lissauer}, {Podolak}, \& {Greenzweig}}]{Pollack:1996}
{Pollack}, J.~B., {Hubickyj}, O., {Bodenheimer}, P., {Lissauer}, J.~J.,
  {Podolak}, M., \& {Greenzweig}, Y. 1996, \icarus, 124, 62

\bibitem[{{Rafikov}(2005)}]{Rafikov:2005}
{Rafikov}, R.~R. 2005, \apjl, 621, L69

\bibitem[{Rhee {et~al.}(2007)Rhee, Song, Zuckerman, \& McElwain}]{Rhee:2007}
Rhee, J.~H., Song, I., Zuckerman, B., \& McElwain, M. 2007, \apj, 660, 1556

\bibitem[{{Riaz} {et~al.}(2006){Riaz}, {Gizis}, \& {Hmiel}}]{Riaz:2006a}
{Riaz}, B., {Gizis}, J.~E., \& {Hmiel}, A. 2006, \apjl, 639, L79

\bibitem[{{Schmidt} {et~al.}(2008){Schmidt}, {Neuh{\"a}user}, {Seifahrt},
  {Vogt}, {Bedalov}, {Helling}, {Witte}, \& {Hauschildt}}]{Schmidt:2008}
{Schmidt}, T.~O.~B., {Neuh{\"a}user}, R., {Seifahrt}, A., {Vogt}, N.,
  {Bedalov}, A., {Helling}, C., {Witte}, S., \& {Hauschildt}, P.~H. 2008, \aap,
  491, 311

\bibitem[{{Steele} {et~al.}(2009){Steele}, {Burleigh}, {Farihi},
  {G{\"a}nsicke}, {Jameson}, {Dobbie}, \& {Barstow}}]{Steele:2009}
{Steele}, P.~R., {Burleigh}, M.~R., {Farihi}, J., {G{\"a}nsicke}, B.~T.,
  {Jameson}, R.~F., {Dobbie}, P.~D., \& {Barstow}, M.~A. 2009, \aap, 500, 1207

\bibitem[{{Subasavage} {et~al.}(2009){Subasavage}, {Jao}, {Henry}, {Bergeron},
  {Dufour}, {Ianna}, {Costa}, \& {M{\'e}ndez}}]{Subasavage:2009}
{Subasavage}, J.~P., {Jao}, W., {Henry}, T.~J., {Bergeron}, P., {Dufour}, P.,
  {Ianna}, P.~A., {Costa}, E., \& {M{\'e}ndez}, R.~A. 2009, \aj, 137, 4547

\bibitem[{{Thalmann} {et~al.}(2009){Thalmann}, {Carson}, {Janson}, {Goto},
  {McElwain}, {Egner}, {Feldt}, {Hashimoto}, {Hayano}, {Henning}, {Hodapp},
  {Kandori}, {Klahr}, {Kudo}, {Kusakabe}, {Mordasini}, {Morino}, {Suto},
  {Suzuki}, \& {Tamura}}]{Thalmann:2009}
{Thalmann}, C., {et~al.} 2009, \apjl, 707, L123

\bibitem[{{Todorov} {et~al.}(2010){Todorov}, {Luhman}, \&
  {McLeod}}]{Todorov:2010}
{Todorov}, K., {Luhman}, K.~L., \& {McLeod}, K.~K. 2010, \apjl, 714, L84

\bibitem[{{van der Bliek} {et~al.}(2004){van der Bliek}, {Norman}, {Blum},
  {Probst}, {Montane}, {Galvez}, {Warner}, {Tighe}, {Delgado}, \&
  {Martinez}}]{van-der-Bliek:2004}
{van der Bliek}, N.~S., {et~al.} 2004, \procspie, 5492, 1582

\bibitem[{{Veras} {et~al.}(2009){Veras}, {Crepp}, \& {Ford}}]{Veras:2009}
{Veras}, D., {Crepp}, J.~R., \& {Ford}, E.~B. 2009, \apj, 696, 1600

\bibitem[{{Wright} {et~al.}(2010){Wright}, {Fakhouri}, {Marcy}, {Han}, {Feng},
  {Johnson}, {Howard}, {Fischer}, {Valenti}, {Anderson}, \&
  {Piskunov}}]{Wright:2010}
{Wright}, J.~T., {et~al.} 2010, arXiv:1012.5676

\bibitem[{{Zuckerman} \& {Becklin}(1992)}]{Zuckerman:1992}
{Zuckerman}, B., \& {Becklin}, E.~E. 1992, \apj, 386, 260

\bibitem[{Zuckerman \& Song(2009)}]{Zuckerman:2009}
Zuckerman, B., \& Song, I. 2009, \aap, 493, 1149

\end{thebibliography}
\end{document}